\documentclass[12pt,a4paper]{article}
\usepackage[utf8x]{inputenc}

\pdfoutput=1

\usepackage{amsmath}
\usepackage{amsfonts}
\usepackage{amssymb}

\usepackage{graphicx}

\usepackage[bookmarks=true,colorlinks=true,citecolor=blue,linkcolor=red,urlcolor=magenta]{hyperref}

\usepackage[authoryear,colon,round,sort&compress]{natbib}

\usepackage{multirow}

\usepackage[table]{xcolor}

\usepackage{geometry} 
\begin{document}

\title{The peculiarities of cross-correlation between two secondary precursors –- radon and magnetic field variations, induced by stress  transfer changes}

\author{V.D.~Rusov$^{1}$\footnote{Corresponding author:  Rusov V.D., E-mail: siiis@te.net.ua}, V.Yu.~Maksymchuk$^2$,  \framebox{R.~Ili\'{c}$^{3,4}$}, V.M.~Pavlovych$^{5}$, R.~Ja\'{c}imovi\'{c}$^4$, \\ V.G.~Bakhmutov$^6$, O.~Kakaev$^1$, V.M.~Vaschenko$^{7,8}$, J.~Skvar\v{c}$^4$, L.~Han\v{z}i\v{c}$^3$, \\
J.~Vaupoti\v{c}$^4$ , M.E.~Beglaryan$^1$, E.P.~Linnik$^1$, S.I.~Kosenko$^1$, D.N.~Saranuk$^1$, \\ V.P.~Smolyar$^1$, A.A.~Gudyma$^8$}

\date{}

\maketitle

\begin{center}
  \textit{$^1$Odessa National Polytechnic University, Shevchenko~av.~1, 65044 Odessa, Ukraine}
  
  \textit{$^2$Carpathian Branch of Institute of Geophysics, National Academy of Science, Naukova~st.~3-b, 79053 Lviv, Ukraine}
  
  \textit{$^3$Faculty of Civil Engineering, University of Maribor, Smetanova~17, 2000 Maribor, Slovenia}
  
  \textit{$^4$J. Stefan Institute, Jamova~39, 1000 Ljubljana, Slovenia}
  
  \textit{$^5$Institute for Nuclear Research, Pr.~Nauki~47, 03028 Kiev, Ukraine}
  
  \textit{$^6$Institute of Geophysics, National Academy of Science, Palladina~av.~32, 03680 Kiev, Ukraine}
  
  \textit{$^7$Ukrainian Antarctic Centre, Tarasa~Schevchenko~Blvd.~16, 01601 Kiev, Ukraine}
  
  \textit{$^8$State Ecological Academy for Postgraduate Education and Management, Uritskogo~str.~35, Kyiv, Ukraine}
  
\end{center}

\abstract{A model of precursor manifestation mechanisms, stimulated by tectonic activity and some peculiarities of observer strategy, whose main task is the effective measurement of precursors in the spatial area of their occurrence on the Earth's daylight, are considered. In particular, the applicability of Dobrovolsky's approximation is analyzed, when an unperturbed medium (characterized by the simple shear state) and the area of tectonic activity (local inhomogeneity caused by the change only of shear modulus) are linearly elastic, and perturbation, in particular, surface displacement is calculated as a difference of the solutions of two independent static problems of the theory of elasticity with the same boundary condition on the surface. Within the framework of this approximation a formula for the spatial distribution (of first component) of magnetic field variations caused by piezomagnetic effect in the case of perturbed regular medium, which is in simple shear state is derived. Cogent arguments in favor of linear dependence between the radon spatial distribution and conditional deformation are obtained.  

Changes in magnetic field strength and radon concentrations were measured along a tectonomagnetic profile of the total length of 11 km in the surroundings of the "Academician Vernadsky" Station on the Antarctic Peninsula (W~64$^{\circ}$16$'$, S~65$^{\circ}$15$'$). Results showed a positive correlation between the annual surface radon concentration and annual changes of magnetic field relative to a base point, and also the good coincidence with theoretical calculation.

}


\section{Introduction}

The information about the concentration of radon isotopes and radon daughters  in the air and soil is actively used all over the world for geophysical purposes \citep{Akerblom1997,Hakl1997,Fleischer1997,Monnin1997,Khan1997,Balcazar1997,Guerra2001,Zmazek2003,Kharatian2002,Majumdar2004}. Monitoring the tectonic activity of the Earth's crust is performed on the basis of several geophysical and chemical methods, including determination of the radon concentration in soil gas and in underground water (e.g. \citealp{Monnin1991}). The applicability of radon is based on the fact that the high temperature of aquifers or geothermal water sources, which are neighbors of natural breaks, promotes the transport of radon upward along existing and/or forming breaks \citep{Segovia1991,Singh1991,Durrani1997}. Measurement of the radon concentration is widely used in seismic testing areas to study active tectonic breaks and earthquake precursors. It is considered that radon is removed by underground waters from cracks in the Earth's crust just made in deformation processes. The close correlation of changes of radon concentration with time in underground waters with movements of the Earth's crust before earthquakes is evidence of that. According to an analogous mechanism, the zones of active tectonic breaks are characterized by anomalous radon concentration.

Study of the tectonic activity in the region of the location of the "Academician Vernadsky" Antarctic station (W~64$^{\circ}$16$'$, S~65$^{\circ}$15$'$) is important because large and deep breaks were revealed near the station.  Furthermore the recent eruptive activity on Deception Island and neovolcanic zone along of Bransfield Strait show the high geodynamic activity on the North of our region \citep{Smellie1988}. The distance between the southern earthquakes in Bransfield Strait rift propagation and Vernadsky station is about 230~km. 

Present geodynamic and seismic tectonic processes, in particular in break zones, lead to changes of mechanical, electrical, magnetic and other properties of rocks. The physical mechanisms of their influence on the variation of magnetic field are due to piezomagnetism and electrokinetic effects. As a result, the temporary changes of geomagnetic field occur with the periods from a few weeks to several years, and the amplitudes from 1 to several tens of nT. These geomagnetic field variations due to the piezomagnetic effect, which are produced by tectonomagnetic anomalies, are the indicators of active geodynamic processes (current movements of the Earth's crust, earthquakes, vulcanization, etc.). They can be revealed by repeated and precise magnetic measurements \citep{Skovorodkin1985}. 

The present work has a twofold aim. First, it is a long-term study of local temporary changes of geomagnetic fields caused by different physical and chemical processes in the Earth's crust. Secondly, it is a search for a correlation between the radon concentration, which reflects the tectonic activity of the Earth's crust, and temporary changes of the abnormal magnetic field. Our results obtained so far are summarized in the present paper.

\section{Theory}

It is well known that the tectonic processes due to regular deformation of the Earth crust are the main cause of existent background deformation field. 

\begin{figure}
\begin{center}
\includegraphics[width=5cm]{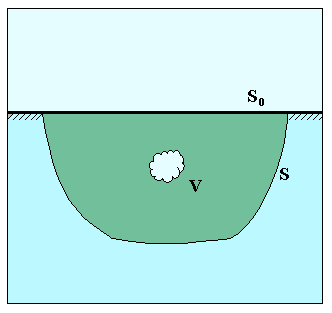}
\caption{The tectonic activity area $V$ in medium.}
\label{fig1}
\end{center}
\end{figure}

We consider that unperturbed medium is at background deformation field, which is supported by corresponding (regional) tectonic processes. We also consider that there is the volume $V$ of tectonic activity, in other words, non-regular deformation volume inside a large unperturbed medium limited by the surface $S = S_0 + S_1$ (Fig.~\ref{fig1}). This local volume has changed properties (heterogeneity), which causes the corresponding perturbations of different geophysical fields in the Earth crust. Traditionally we call such perturbations "precursors". The main, or more exactly, primary precursor of tectonic activity is a so-called mechanical precursor, i.e. medium deformation, exceeding background deformation. All the other precursors including those discussed in this paper -- anomalies of radon concentration and magnetic variations -- are the secondary with respect to Earth crust deformations.

A model describing mechanisms of precursor manifestations stimulated by tectonic activity and some peculiarities of observation strategy for effective precursor measurements in spatial zone of their manifestation on the daylight of the Earth is presented below.

\subsection{The distributions of deformations and radon concentrations along the Earth day surface}
\label{sec2.1}

We define a zone of precursor activity manifestation as a part of daylight, which contains the epicenter of tectonic stress and is limited by a line where precursor perturbations are of the same magnitude as the background variations. Taking into account this definition let us consider the construction of deformation distribution on daylight for a typical case of tectonic activity. Note that the main idea of such problem solution (excluding some modification of process phenomenology) is completely based on Dobrovolsky theory of seismic focus evolution \citep{Dobrovolsky1984}. 

As it is stated above, the Earth crust moves with approximately constant (on regional scale) strain rate, which maintains a constant (in mentioned sense) level of shearing stress against a background of almost hydrostatic stress field due to gravitation. 

A viscous-elastic medium is the simplest model of continuous medium, which reflects the stated properties. However the high effective viscosity of the Earth core (which is 10$^{20}$-10$^{22}$~Pa/s according to different estimates), relatively short duration of investigated processes (1~-~10~years) and necessity to study the difference of states rather than the characteristics of absolute state make it possible to introduce the following simplification. We consider an approximation, when the unperturbed medium and the volume of tectonic activity (local heterogeneity) are linearly elastic and the perturbation is calculated as the difference of the solutions of two separate static problems of theory of elasticity with the same boundary conditions on surface (Fig.~\ref{fig1}). Equal boundary conditions stipulated by the fact that the energy of tectonic processes considerably exceeds the increment of the energy due to the formation of the area of increased tectonic activity and therefore the appearance of local heterogeneity cannot distort the regular tectonic processes far off its hypocenter.

Let us consider half-space $x_3 \geqslant 0$, which contains homogeneous and isotropic medium. The problem is set in rectangular coordinates $x_i$ (the comma of the lower index denotes differentiation by corresponding spatial coordinate and the recurring indexes cause summation). In the general case, at small $\alpha$, which characterizes the weight of perturbed part of elastic modulus, the perturbed displacements $w_r$ of such medium are described by Dobrovolsky's formula \citeyearpar{Dobrovolsky1984}

\begin{equation}
w_{r} (x) = - \alpha \left[ \frac{\sigma^{0}_{ll}}{3} \frac{K'}{K^{0}} \delta_{i}^{j} \frac{\mu '}{\mu^{0}} \left( \sigma^{0}_{ij} - \frac{\sigma^{0} _{ll}}{3} \delta_{i}^{j} \right) \right] \int \int \limits_{V} \int \upsilon^{r}_{i,j} (\xi, x) d \upsilon_{\xi},
\label{eq1}
\end{equation}

\noindent where 

\begin{equation}
\sigma_{ij}^0 = c_{ijkl}^0 \varepsilon_{kl}^0
\label{eq2}
\end{equation}

\noindent is unperturbed stress equal to the product of unperturbed elastic modulus  and deformation value $\varepsilon_{kl}^0$, $K'$ and $K^0$ are the bulk elastic modules, $\mu '$ and $\mu^0$ are shear modules in perturbed and unperturbed state, respectively, $\delta_i ^j$ are Kronecker indices, $ \upsilon^{r}_{i,j} (x, \xi)$ is Green function (tensor), $\xi$ are the coordinates of the origin of unit point force and $V$ is the volume of the perturbed area of investigated medium.

According to Dobrovolsky, we suppose that only shear modulus is changed in the area of increased tectonic activity $V$, i.e. $K' = 0$, $\mu ' = \mu^0 = \mu$, and regular (unperturbed) state is state of pure shear with stress

\begin{equation}
\sigma_	{12}^{0} = \sigma_	{21}^{0} = \tau, ~~~ for~another~\sigma_{ij}^{0} = 0.
\label{eq3}
\end{equation}

Then Eq.~(\ref{eq1}) with allowance of Eq.~(\ref{eq3}) has the following form

\begin{equation}
w_{r} (x) = - \alpha r \int \int \limits_{V} \int \left[ \upsilon^{r}_{1,2} (\xi, x) + \upsilon^{r}_{2,1} (\xi, x) \right] d \upsilon_{\xi},
\label{eq4}
\end{equation}

At the same time, when surface displacements are calculated the Green function  is the solution of the problem of point unit force applied on surface, i.e. so-called Boussinesq-Cerutti tensor \citep{Novatsky1975}. For the r$^{\text{th}}$ unit power  applied in the point ($\xi_1$, $\xi_2$, 0) we have

\begin{align}
\upsilon^{r}_{i} (x, \xi) & = \frac{1}{4 \pi \mu} \left[ \left( \frac{1}{R} + \frac{1-2\nu}{R+x_3} \right) \delta_{i}^{r} + \left( \frac{1}{R} - \frac{1-2\nu}{R (R+x_3)^2} \right) \left( x_i - \xi _i \right) \left( x_r - \xi _r \right) + \right. \nonumber \\ 
& + \frac{1-2\nu}{R (R+x_3)} \left( \left( x_r - \xi_r \right) \delta_i ^3 - \left( x-i - \xi_i \right) \delta_r ^3 \right) + \nonumber \\
& + \frac{(1-2\nu) x_3}{R (R+x_3)^2} \left( \left( x_r - \xi_r \right) \delta_i ^3 + \left( x-i - \xi_i \right) \delta_r ^3 \right) + \nonumber \\
& + \left. \left( \frac{(1-2\nu)}{R} - \frac{1-2\nu}{R+x_3} - \frac{x_3^2 (1 - 2\nu)}{R+x_3} \right) \delta_{i}^3 \delta_{r}^3 \right],
\label{eq5}
\end{align}

\noindent where $\nu$ is Poisson's ratio, and

\begin{equation}
R = \sqrt{\left( \left( x_1 - \xi_1 \right)^2 + \left( x_2 - \xi_2 \right)^2 + x_3 ^2\right)}
\label{eq6}
\end{equation}

It turns out that the perturbed displacements, Eq.~(\ref{eq4}), with allowance of Eq.~(\ref{eq5}) can be calculated in elementary functions in case, when area $V$ is a parallelepiped \citep{Dobrovolsky1976}. It is also shown ibidem that for areas having more complex form or for points located outside of area $V$, sufficiently far from this area, it is convenient to calculate $w_r$ approximately, substituting the integrand by its value in the some internal point of volume, which coincides with the center of area $V$, when it is symmetric (monopole approximation). Note that the error of such approximate calculation is less than a few percents even in epicenter zone in comparison with exact solution  \citep{Dobrovolsky1976}.

Let us place the centre of area $V$ at point $(0, 0, H)$ and introduce the following designations

\begin{equation}
x = x_1, ~~ y = x_2, ~~ r = \sqrt{x^2 + y^2 + H^2}
\label{eq7}
\end{equation}

Then the approximate calculation of integral in Eq.~(\ref{eq4}) gives displacements in the following form

\begin{equation}
w_1 = - \frac{\alpha V \tau}{2 \pi \mu} \frac{y}{r^2} \left[ \frac{3 x^2}{r^2} + \frac{1 - 2\nu}{(r + H)^2} \left( \frac{r^2 - x^2}{r} - \frac{2 x^2}{r + H} \right) \right],
\label{eq8}
\end{equation}

\begin{equation}
w_2 = - \frac{\alpha V \tau}{2 \pi \mu} \frac{x}{r^2} \left[ \frac{3 y^2}{r^2} + \frac{1 - 2\nu}{(r + H)^2} \left( \frac{r^2 - y^2}{r} - \frac{2 y^2}{r + H} \right) \right],
\label{eq9}
\end{equation}

\begin{equation}
w_3 = - \frac{\alpha V \tau}{2 \pi \mu} \frac{xy}{r^2} \left[ \frac{3 H}{r^2} + \frac{(1 - 2\nu) (2r + H)}{(r + H)^2}\right],
\label{eq10}
\end{equation}

Further, using Eqs.~(\ref{eq8})-(\ref{eq10}) the deformation tensor components are calculated

\begin{equation}
\varepsilon_{xx} = \frac{\alpha V \tau}{2 \pi \mu} \frac{3xy}{r^3} \left[ \frac{2}{r^2} - \frac{5x^2}{r^4} + \frac{1 - 2\nu}{(r + H)^3} \left( \frac{2x^2}{r + H} - \frac{\left( y^2 + H^2 \right) \left( 3r + H \right)}{r^2} \right) \right],
\label{eq11}
\end{equation}

\begin{equation}
\varepsilon_{yy} = \frac{\alpha V \tau}{2 \pi \mu} \frac{3xy}{r^3} \left[ \frac{2}{r^2} - \frac{5y^2}{r^4} + \frac{1 - 2\nu}{(r + H)^3} \left( \frac{2y^2}{r + H} - \frac{\left( x^2 + H^2 \right) \left( 3r + H \right)}{r^2} \right) \right],
\label{eq12}
\end{equation}

\begin{align}
\varepsilon_{yy} & = \frac{\alpha V \tau}{2 \pi \mu} \frac{1}{r^3} \left[ \frac{3(r^2 - H^2)}{2r^2} - \frac{15 x^2 y^2}{r^4} + \frac{1 - 2\nu}{2(r + H)^2} \times \right. \nonumber \\
& \left. \times \left( 6r^2 - r H^2 + \frac{4 x^2 y^2 (5r + 2H)}{r (r + H) ^2} - \frac{\left( r^4 + x^4  + y^4 - H^2 \right) \left( 5r + 3H \right)}{r^2 (r + H)} \right) \right],
\label{eq13}
\end{align}

In order to present the strain distribution on daylight in a convenient or universal form we introduce the dimensionless variables

\begin{equation}
\xi = x / \rho, ~~~ \eta = y / \rho, ~~~ h = H / \rho
\label{eq14}
\end{equation}

\noindent and conditional deformation

\begin{equation}
\omega _{ij} = \frac{3 \mu}{2 \alpha \tau} \varepsilon _{ij},
\label{eq15}
\end{equation}

\noindent where $\rho$ is the average radius of area $V$, i.e. the sphere radius of same volume.

\begin{figure}
\begin{center}
\includegraphics[width=10cm]{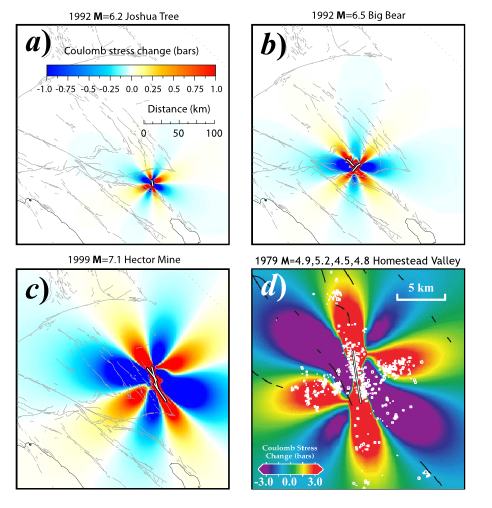}
\caption{Coulomb stress changes associated with (a-b-c) the three M$\geqslant$6 earthquakes in the same seismic region (from \citet{Toda2005}) and (d)  the Homestead Valley earthquakes sequence (M=4.9, 5.2, 4.5, and 4.8) (from \citet{King1994}). Red (positive) indicates that optimally oriented faults are stressed more towards failure and blue/purple (negative) indicates that failure is inhibited. White circles are observed aftershocks.}
\label{fig1a}
\end{center}
\end{figure}

A question arises here of whether the application of the modified Dobrovolsky \citeyearpar{Dobrovolsky1984} theory results is justified or not. Strange as it seems, the effective verification of Dobrovolsky theory results is possible. It turns out to be not too difficult to show that if the length and depth of the fault zone are comparable, the displacements distribution results, obtained within the so called Coulomb stress change (e.g. Fig.~\ref{fig1a}) analysis \citep{King1994,Stein1999,Cocco2002,Kanamori2004,Toda2005} and the seismic focus formation theory \citep{Dobrovolsky1984} must be the same.

It is interesting to note in this regard that, as it will be shown in Section \ref{sec4}, the displacements distributions, obtained withing the mentioned approaches, really coincide. It is a very important result and here is why. It is known \citep{King1994} that the strategy for studying aftershocks is to map the calculated Coulomb stress change based on the observed slip during an earthquake and compare the resulting field with the observed aftershock distribution (red region on the Fig.~\ref{fig1a}d). And vice versa, the low-strain (e.g. blue/purple region on the Fig.~\ref{fig1a}) regions play the role of highly attractive targets for a soil radon. In other words, since in some areas the strain is extensional while in the others is is compressional, the soil radon flows, which primarily originate from much deeper than the fault zone, "choose" their route through the extensional regions because of a natural pressure gradient.

That is why in addition to registration of soil radon, a study of its spatial correlation with extensional areas appearing in the near-field of the seismic focus was one of the goals for present paper. The existence of such correlation may, on the one hand, be a serious reason to consider the soil radon a secondary precursor of the earthquake, and on the other hand, be a key to understanding of the anomalous radon spatial distribution pattern. Anomalous radon behavior usually manifests itself in its sudden "disappearance" just before the ongoing earthquake. As it will be shown below, it does not actually disappear, but rather is being redistributed according to the spatial extensional areas distribution (e.g. blue/purple region on the Fig.~\ref{fig1a}). Hence such an anomaly is a mere consequence of the detectors location within the compressional areas (red region on the Fig.~\ref{fig1a}d) which naturally explains the known mystery of the radon sudden disappearance before the earthquake.

Moreover, the spatial correlation of the soil radon with extensional areas, which is a definite indicator of any secondary precursor, may be amplified by a simultaneous correlation with some other secondary precursor. For that purpose let us consider the magnetic field variations induced by a process of the seismic focus formation itself.

\subsection{The distributions of deformations and magnetic field along the Earth day surface}
\label{sec2.2}


Variations in electromagnetic field following large earthquakes are frequently reported \citep{Eleman1965,Honkura2000,Iyemori1996,Taira2009,Takla2012}. One of the candidate mechanisms that generate coseismic variations in the magnetic field is the piezomagnetic effect. The piezomagnetic effect generates changes in magnetization in the Earth's crust under the application of mechanical stress. When a change of the mechanical stress field causes the corresponding change in magnetization $\Delta I$ in every point of a continuous medium, the piezomagnetic effect is explained by the fact that a change in the magnetic moment of every volume element changes the magnetic induction caused by this elementary magnetic dipole variation.

At the same time, it is known, that changes in the magnetic field due to the piezomagnetic effect, which are reffered to as the piezomagnetic field, have been calculated for a variety of elastic models (e.g. \citet{Dobrovolsky1976,Dobrovolsky1984,Sasai1991,Sasai1994,Sasai2001,Utsugi1999,Okubo2004}). It has been established historically that in these earlier studies, only permanent displacements of elastic materials were considered. The fact that to investigate variations in the magnetic field during the propagation of seismic waves, calculations should be extended in such a way that time-dependent stress fields are properly treated, had been frequently ignored. In other words the contradictions, according to \citep{Yamazaki2011a} lies in the fact that in the Stokes's solution of the theory of elasticity, permanent displacements are represented by near-field terms, whereas seismic wave are mainly represented by far-field term \citep{Aki2002}. This contradiction becomes more apparent if one takes into account that the consideration of elastic waves corresponding to the far-field terms is essential when considering the piezomagnetic field observed at sites located far (i.e. several tens of kilometers) from the seismic source \citep{Yamazaki2011a}. 

A substantial progress in studying the piezomagnetic effect was achieved recently in papers by Yamazaki \citeyearpar{Yamazaki2009,Yamazaki2011a,Yamazaki2011b}, where the mentioned problems are effectively overcome. It is also shown there that, on the one hand, for the homogeneous (regarding the initial magnetization) medium with dominant period of Rayleigh waves ($20-30 s$) the magnitudes of variations due to the piezomagnetic effect are very weak and hardly reaches $0.1~nT$, which is close to the limit of detectability \citep{Yamazaki2011a}, and on the other hand, the amplitudes of the piezomagnetic signals arising from non-uniformly magnetized crust are as large as $0.5~nT$ \citep{Yamazaki2011b}. The latter indicates that the piezomagnetic field may be a plausible mechanism of generating co-seismic changes in the magnetic field with detectable amplitudes for large earthquakes, provided that the observation site is located near the magnetization boundaries \citep{Yamazaki2011b}.

So it may be noted here that the systematic study of the phenomenon under question shows that under  increased tectonic activity, including the earthquake preparation processes, the magnetic field variations appear that in certain cases (strongly non-uniformly magnetized crust and magnetic field observation in the near-field of earthquake) are really determined by the stated processes and may be caused by the piezomagnetic effect, which had been described for the first time in the papers by \citep{Breiner1964,Stacey1972}. Furthermore, below we shall use the following expression for the piezomagnetic effect description, obtained in the paper \citep{Johnston1973,Johnston1978}, but presented in the coordinate system coinciding with the principal axes of stress tensor.

\begin{equation}
\Delta I_i = C I_i \left( \frac{\sigma_ j + \sigma_k}{2} - \sigma_i \right),
\label{eq16}
\end{equation}

\noindent where $\sigma_ i$, $\sigma_ j$ and $\sigma_ k$ are principal stresses (all $i$, $j$ and $k$ are different) and $C$ is piezomagnetic coefficient. 

Although all terms in Eq.~(\ref{eq16}) are tensors, the equation is not of invariant (relative to rotation) tensor form. However, as \citet{Dobrovolsky1984} showed that Eq.~(\ref{eq16}) may be written in a more handy form. Really, there is an obvious equality

\begin{equation}
\frac{\sigma_ j + \sigma_k}{2} - \sigma_i = - \frac{3}{2} \left( \sigma_i - \frac{\sigma_i + \sigma_j + \sigma_k}{3} \right).
\label{eq17}
\end{equation}

Here the expression in brackets is the deviator of stress tensor in principal axes. If deviator components are expressed as $s_{ij}$, the Eq.~(\ref{eq16}) has the following form

\begin{equation}
\Delta I_i = - \frac{3}{2} C I_j s_{ij},
\label{eq18}
\end{equation}

\noindent where there is summation by recurring indexes. It is obvious that, by virtue of rules of tensor algebra, Eq.~(\ref{eq18}) in such form is invariant relative to rotations and therefore is valid in arbitrary coordinates. 

Now let us calculate the distribution of magnetic field regional variations $\Delta F_r$ determined by difference between the total vector of magnetic induction and the background induction $\vec{B}_E$, characterized exceptionally by basic terrestrial magnetic field

\begin{equation}
\Delta F_r = \left \vert \vec{B} _E + \Delta F_r \right \vert - \left \vert \vec{B} _E \right \vert.
\label{eq19}
\end{equation}

When $\left( \left( \Delta F_r \right)^2 / B_E \right) \ll 1$ the Eq.~(\ref{eq19}), with adequate accuracy, may be presented as

\begin{equation}
\Delta F_r = \frac{\vec{B}_E}{B_E} \Delta \vec{B} _r = \vec{b} \cdot \Delta \vec{B} _r,
\label{eq20}
\end{equation}

\noindent where $\vec{b}$ is unit vector of basic terrestrial magnetic field.

On the other hand, it is well known that magnetic induction may be unambiguously expressed by vector potential $\vec{A}$ in the following way

\begin{equation}
\vec{B} = rot \vec{A}, ~~~ \text{if}~ div \vec{A} = 0,
\label{eq21}
\end{equation}

\noindent where the equality to zero of vector potential divergence ensures the potential calibration. 

From magnetism physics it is also known that vector potential of magnetic dipole with magnetic moment $\vec{m}$ has form

\begin{equation}
\vec{A} = k \frac{\vec{m} \times \vec{R}}{R^3},
\label{eq22}
\end{equation}

\noindent where $k$ is the coefficient dependent on used system of units, and $\vec{R}$  is the radius vector directed from dipole point to the point of observation.

If in volume element $dv$ the change of magnetization occurred, it would cause magnetic momentum variation of value  and the corresponding variation of vector potential

\begin{equation}
d \vec{A} = k \frac{\Delta \vec{I} \times \vec{R}}{R^3} dv.
\label{eq23}
\end{equation}

Integration of the Eq.~(\ref{eq23}) over the total Earth's core layer from daylight to Curie isotherm gives the total variation of vector potential. It reduces to the following form of Eq.~(\ref{eq21})

\begin{equation}
\Delta \vec{B} = \Delta \vec{F} = k \cdot rot \int \int \int \frac{\Delta \vec{I} \times \vec{R}}{R^3} dv = k \nabla \times \int \int \int \frac{\Delta \vec{I} \times \vec{R}}{R^3} dv,
\label{eq24}
\end{equation}

\noindent where $\nabla$ is nabla operator.

The insertion of expression for magnetization, Eq.~(\ref{eq18}) in Eq.~(\ref{eq24}), taking into account scalar product (Eq.~(\ref{eq20})) and considering also the Levi-Civita symbols $e_{ijk}$, the final solution for $\Delta B$ has the form

\begin{equation}
\Delta F = - \frac{3}{2} k C b_p e_{mkp}  \nabla _m \int \int \int \frac{e_{ijk} I_l s_{il}' R_j}{R^3} dv_{\xi},
\label{eq25}
\end{equation}

\noindent where $\nabla _m  = \partial / \partial x_m$; $R_j = x_j - \xi_j$ ($x_j$ are the coordinates of the point of observation, $\xi_j$ is the integration variable) and $s_{il}'$ is the stress tensor deviator.

Further we shall use the expression of stress increment obtained in \citep{Dobrovolsky1984}

\begin{equation}
\sigma _{ij} ' = \alpha \sigma _{ij} ^0 \delta _V + \sigma _{ij} '',
\label{eq26}
\end{equation}

\noindent where $\sigma _{ij} '' = c_{ijkl}^0 w_{kl}$ ($c_{ijkl}^0$ is the elastic modulus of unperturbed medium), $w_{kl} = u_{kl} - u_{kl}^0$, i.e. the difference of perturbed and unperturbed displacements and $\delta_V$ is the characteristic function of volume $V$. In this case Eq.~(\ref{eq26}) in corresponding deviators looks like

\begin{equation}
s_{ij} ' = \alpha s_{ij} ^0 \delta _V + s_{ij} ''.
\label{eq27}
\end{equation}

It follows from Eq.~(\ref{eq27}) that Eq.~(\ref{eq25}) divides into two parts

\begin{equation}
\Delta F_r = \Delta F_{r1} + \Delta F_{r2}.
\label{eq28}
\end{equation}

Let us investigate both components separately. In the first item,

\begin{equation}
\Delta F_{r1} = - \frac{3}{2} k C b_p e_{mkp} \nabla \int \int \int \frac{e_{ijk} I_l \alpha s_{il} ^0 R_j}{R^3} dv,
\label{eq29}
\end{equation}

\noindent the integration takes place only over the volume $V$, therefore the assumption $I = const$ is appropriate for the estimation. Since it is possible to consider that $\vec{I} \| \vec{B}_E$ for the isotropic domain $V$, we have $I_l = I b_l$. As a result, Eq.~(\ref{eq29}) with allowance for  $s_{ij}^0 = const$ takes form

\begin{equation}
\Delta F_{r1} = - \frac{3}{2} k C \alpha I  b_p e_{mkp} \nabla e_{ijk} b_l s_{il} ^0 \int \int \int \frac{R_j}{R^3} dv,
\label{eq30}
\end{equation}

Since we are interested in distribution of magnetic field variations on the Earth's daylight, calculating integral in Eq.~(\ref{eq30}) in monopole approximation, we obtain an approximation for Eq.~(\ref{eq30})

\begin{equation}
\Delta F_{r1} = - \frac{3}{2} k C \alpha I  b_p e_{mkp} \nabla e_{ijk} b_l s_{il} ^0 \frac{r_j}{r^3},
\label{eq31}
\end{equation}

\noindent where $r_j = x_j \xi_j '$ and $\xi_j '$ are coordinates of some point within the area $V$.

On the basis of exact calculation of Eq.~(\ref{eq30}) for a sphere, which tangents to half-space boundary at the point of contact $x_j$, \citet{Dobrovolsky1984} showed that monopole approximation~(\ref{eq31}) is satisfactory, if $\xi_j '$ is located closer to daylight approximately on half radius relative to the center of area $V$. As a result we take in Eq.~(\ref{eq31})

\begin{equation}
\xi _j ' = \xi _j ^0 - \frac{1}{2} \delta _{3j} \rho,
\label{eq32}
\end{equation}

\noindent where $\xi_j ^0$ are the coordinates of area $V$ center and $\rho$ is the average radius of area $V$, i.e. the radius of a sphere with equal volume.

The use of Eq.~(\ref{eq32}) is justified only for calculations in epicenter zone, whereas the admission $\xi_j ' = \xi_j ^0$ already on the boundary gives the result differing from those calculated by Eq.~(\ref{eq32}) up to 25\%.

Since

\begin{equation}
\nabla _m \frac{r_j}{r^3} = - \frac{1}{r^3} \left[ \frac{3 (x_j - \xi_j) (x_m - \xi_m)}{r^2} - \delta _m ^j \right] = - \frac{1}{r^3} q_{jm},
\label{eq33}
\end{equation}

\noindent (where $q_{jm} = q_{mj}$ and $q_{jj} = 0$) Eq.~(\ref{eq31}) after simplifications takes the form 

\begin{equation}
\Delta _{r1} = \frac{3}{2} \frac{k C V \alpha I}{r^3} b_m b_j s _{ij} ^0 q_{im}.
\label{eq34}
\end{equation}

As in our case a regular (unperturbed state) is considered as simple shear state with stress Eq.~(\ref{eq3}) we obtain from Eq.~(\ref{eq34}) and, consequently, from Eq.~(\ref{eq29})

\begin{equation}
\Delta F_{r1} = - \frac{3}{2} \frac{k C V \alpha I}{r^3} \left[ \left( b_1 ^2 + b_2 ^2 \right) q_{12} + b_2 b_3 q_{13} + b_1 b_3 q_{23} \right],
\label{eq35}
\end{equation}

\noindent where $b_1 = \sin \theta _1$ and $b_2 = \cos \theta_2$, $b_3 = \sin \theta_2$. Here $\theta_1$ and $\theta_2$ are the angles determining the inclination and declination of the vector of the main terrestrial magnetic field in given region). 

Thus, the Eq.~(\ref{eq35}) is improved \citet{Dobrovolsky1984} formula for the spatial distribution of magnetic field variations (first component) in the case of perturbation of regular medium, which is the simple shear state. Here the multiplicative coefficient before the bracket contains the information about the scalar magnitudes of magnetic characteristics, stresses in medium and dependence on distance. The expression in the brackets, in general, describes the direction distribution and therefore is called direction function. The further concretisation of Eq.~(\ref{eq35}) for $\Delta F_{r1}$ connected with magnetic description of the investigated Earth's region and the orientation of tectonic stresses are considered in the section 3.

\section[Experimental]{Experimental\footnote{The experiments were carried out during the 7$^{\text{th}}$, 8$^{\text{th}}$, 9$^{\text{th}}$ and 10$^{\text{th}}$ Ukrainian Antarctic Expeditions in March 2002, March 2003, March 2004 and March 2005, respectively}}

\begin{figure}
\begin{center}
\includegraphics[width=0.7\linewidth]{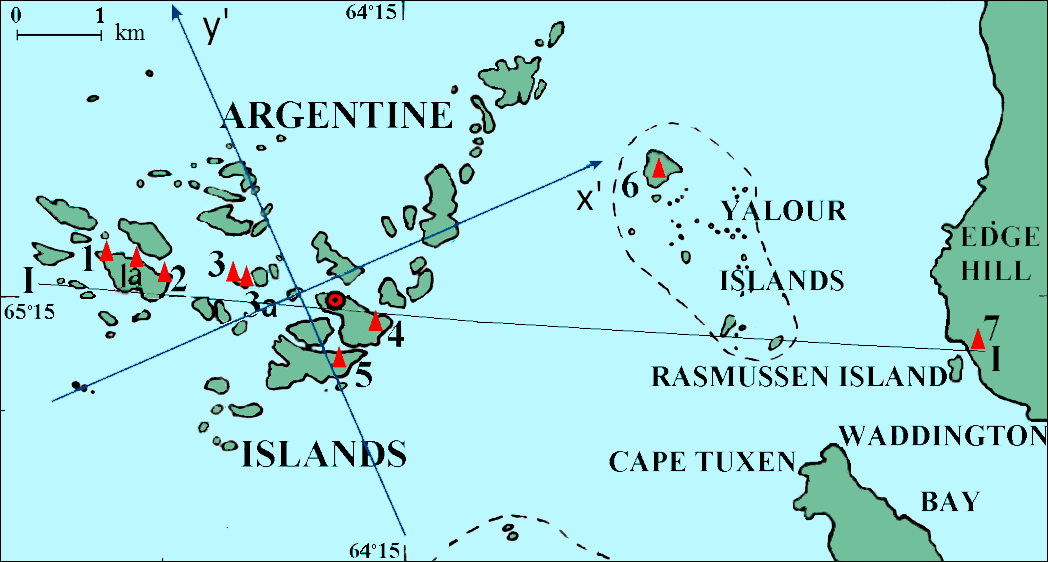}
\caption{Location of geomagnetic and radon measurements in the region "Academician Vernadsky" Antarctic Station. BP - base points for magnetic and radon observation "Argentine Islands"; I - I - latitudinal profile. Axis $X$ is line of maximal tangential stresses in given region.}
\label{fig2}
\end{center}
\end{figure}

The tectonomagnetic investigations in the vicinity of the "Academician Vernadsky" station began in 1998 along the sublatitudinal profile Barchans Islands – Rasmussen point (Fig.~\ref{fig2}). In 2001 the first data about geomagnetic field $\Delta F$ time changes were reported in \citep{Maksymchuk2002}. The next cycle of observations was carried out in 2002 and extended by addition of new tectonomagnetic sites of observation. The fourth, fifth and sixth measurements were performed in 2003, 2004 and 2005, respectively \citep{Maksymchuk2004,Maksymchuk2006}. Detailed results of these investigation will be published elsewere \citep{Maksymchuk2006}.

The measurement of radon concentrations was performed at several sites close to the sites of tectonomagnetic observations along the Barchans-Rasmussen profile. These measurements were started during the 7$^{\text{th}}$ season Ukrainian Antarctic Expedition in 2002. It should be mentioned that in 2003 the $Rn$ laboratory was set up close to the base point (BP), where continuous $Rn$ measurements with an active $Rn$ monitor were performed \citep{Ilic2005}. In 2004 an improved $Rn$ monitoring procedure for etched track detectors was introduced and 3 additional continuous $Rn$ devices were set up to extend $Rn$ studies to other topics \citep{Rusov2012,Ilic2005}.

\subsection{Site}

The "Academician Vernadsky" Antarctic station (the former UK station "Faraday") is located on Galindez Island (Archipelago of the Argentine Islands), part of the West Pacific Shelf of the Antarctic Peninsula (Fig.~\ref{fig2}). The Antarctic Peninsula with its adjacent Islands forms a part of the West Antarctic folded system, which extends along the Pacific Coast of Western Antarctica. The origin of the Antarctic Peninsula is located in the southern part of the jointing zone of two large-scale geological structures – the Pacific and Gondvana segments – of the Earth. It is a result of long-time interaction of tectonic processes, which are peculiar to each of these planetary segments and cause their main peculiarities as a whole. The deep trench of the Wedell Sea and its outlying western and southwestern shelf belong to the Gondvana segment. At the present stage of evolution, the deep ocean trench and marine shelf area, as well as the Antarctic Peninsula with its adjacent islands and lands, belong to the Pacific Segment. 

The evidence of the geodynamic activity of the region is neovolcanic zones and seismic activity in Bransfield Strait. The South Shetland Trench, which lies to  N-W part of the South Shetland Islands, is the sole  remnant zone of the subduction zone that existed along the entire Pacific margin of West Antarctica during the Mesozoic and Cenozoic \citep{Baker1982}. Most of the subduction zone was progressively eliminated as the Phoenix-Antarctic spreading center was  subducted beneath the Antarctic Peninsula. Slow subduction of the former Phoenix Plate appears to continue today within what is essentially a single Antarctic plate, owing to slab-pull forces and/or trench rollback. The seismic, multibeam bathymetry and geodetic surveys data indicates that Bransfield Strait is undergoing active extension with the rate of $\sim$10 mm/yr in direction orthogonal to the strait's strike \citep{Christenson2003}. The modern eruptive activity observed on Deception Island extends along this neovolcanic zone.

 The most part of the western Antarctic Peninsula is represented by Meso-Cenozoic intrusive complex and Jurassic-Cretaceous calk-alkaline AP Volcanic Group. The oldest rocks in Argentine Islands Archipelago are laves and pyroclastic members of the Upper Jurassic Volcanic Group which have been intruded by pre-Andean dykes and sills. The Andean Intrusive Suite has metamorphosed and metasomatized the volcanic and hypabyssal rocks. There is a later dyke phase which cuts both of volcanic and plutonic rocks. Finally, there are a few late comparatively unaltered Tertiary dykes.  Whereas the Argentine Islands are composed entirely of the volcanic rocks (andesite laves and dacite pyroclastic rocks) the Andean Intrusive Suite rocks represented by granodiorites occur in the western part of archipelago.
 
The tectonomagnetic points of observation were selected mainly in the Argentine Islands Archipelago (Fig.~\ref{fig2}, Table~\ref{tab1}). Two tectonomagnetic profiles could be chosen here: a latitudinal profile (Barchans Islands – Rasmussen point), and a meridian profile (Berselot Island – Booth Island). The total length of latitudinal profile is 11~km (having 7 observation points, denoted by I – I in Fig.~\ref{fig2}). The profile from W to E crosses the volcanic and plutonic rocks with the probably fault zone between them, the fault between Argentine Islands Archipelago and Antarctic Peninsula in Penola Strait which well traced in bathymetry and reaches the Antarctic Peninsula on the East. 

\begin{table}
\begin{center}
\caption{Sites of measurements}
\label{tab1}
\begin{tabular}{|c|c|c|}
  \hline
  Location Number (n) & Location Name & Coordinates \\
  \hline
 1 & Barhans Islands west & S 65$^{\circ}$14.362$'$ W 64$^{\circ}$19.023$'$ \\
  \hline
 1a & Barhans Islands central & S 65$^{\circ}$14.410$'$ W 64$^{\circ}$18.447$'$ \\
  \hline
 2 & Barhans Islands east & S 65$^{\circ}$14.544$'$ W 64$^{\circ}$17.862$'$ \\
  \hline
 3 & Three Little Pigs west & S 65$^{\circ}$14.617$'$ W 64$^{\circ}$16.855$'$ \\
  \hline
 3a & Three Little Pigs west & S 65$^{\circ}$14.630$'$ W 64$^{\circ}$16.761$'$ \\
  \hline
 BP & Galindez Island Basic Point (BP) & S 65$^{\circ}$15$'$ W 64$^{\circ}$16$'$ \\
  \hline
 4 & Galindez Island Penguin Point & S 65$^{\circ}$14.919$'$ W 64$^{\circ}$14.332 \\
  \hline
 5 & Squa Island & S 65$^{\circ}$15.225$'$ W 64$^{\circ}$15.029$'$ \\
  \hline
 6 & Yalour Island & S 65$^{\circ}$14.035$'$ W 64$^{\circ}$09.029$'$ \\
  \hline
 7 & Rasmussen Cape & S 65$^{\circ}$14.852$'$ W 64$^{\circ}$19.023$'$ \\
 \hline
\end{tabular}
\end{center}
\end{table}

\subsection{Monitoring of tectonomagnetic anomalies}

Tectonomagnetic anomalies, defined as abnormal temporary changes of geomagnetic field due to physical and chemical phenomena in the Earth's crust and upper mantle, are results of changes of seismic, volcanic and other geodynamic processes. The study of such changes of magnetic field is connected with some difficulties since their amplitude does not exceed several tens of nT and the spectrum overlaps with the spectrum of variations of internal (with the duration over 11 years) and external origin (ionosphere and magnetosphere with a duration from seconds to several years).

Among numerous physical mechanisms which can lead to the appearance of tectonomagnetic anomalies, piezomagnetic and electrokinetic effects are the most predominant. Both of these effects are connected with changes in the stressed and deformed state of the geological medium, which lead to changes of magnetic properties of the rocks and variation of the electrical current. 

The piezomagnetic effect is manifested by a change of the magnitude and space distribution of the magnetization vector due to mechanical loading in the range of elastic deformation. Electrokinetic phenomena appear in an electrically neutral system as a whole and consist of the displacement of one phase (liquid or solid) by another under the influence of an external electric field, or in the appearance of an electrical current due to phase displacement under mechanical stress. 

Theoretical calculation of the anomalous effects for both mechnisms gave approximate estimates, which showed that the changes of magnetic field $\Delta F$ can vary in the range from 0.1 to~10~nT with pressure variations of 1~-~100 bars/year \citep{Maksymchuk2001}. Besides this, the anomalous effects strongly depend on the depth and shape of the heterogeneities, element contents, magnetic and electric properties, etc. It should be noted that the pressure, even in strong earthquakes, does not exceed 100~bar. 

The classical method for detection of tectonomagnetic anomalies is based on measurements of the Earth's magnetic field at a constant set of the observation points in definite subsequent time intervals. Usually, for technical reasons, the absolute value of the Earth's magnetic field vector F is measured. In order to avoid the influence of external field changes, the work was performed according to the scheme of synchronous differential measurements of the field $F_{BP}$ at a base point (BP) and the field $F_n$ at given ordinary point (n) of observation. The field difference $\Delta F = F_n - F_{BP}$ is independent of the external field influence for small distances (10~-~30~km) between the base and ordinary points. Thus the  parameter sought is $\Delta \Delta F$ i.e. the change of $\Delta F$ during the period between the cycles of observations defined as

\begin{equation}
\Delta \Delta F = \Delta F_Y - \Delta F_{Y-1}
\label{eq36}
\end{equation}

\noindent where $\Delta F_Y$ and $\Delta F_{Y - 1}$ are the measured values of $\Delta F$ in the year $Y$ and $Y-1$ respectively. 

The interval between the cycles of measurements is determined by the problem under study and the parameters of the phenomena, and can amount to days, months or years. Taking into account the climatic conditions of Antarctica (Archipelago of the Argentine Islands), it is most convenient to make measurements once a year, i.e. the studied tectonomagnetic effects have a period of about one year. 

As the base point of observation we used the Argentine Islands (AIA) magnetic observatory, where $F$ was recorded by MB-01 proton magnetometer with a sensitivity of $\pm 0.1 ~nT$. Measurements of the field at the set of ordinary points were performed by  MPP-203 proton magnetometer. The standard deviation of the field determination was $0.5-1 ~nT$. Results of geomagnetic field measurements are summarized in Table~\ref{tab2}.

\begin{table}
\begin{center}
\caption{Results of geomagnetic field and radon concentration measurements at the tectonomagnetic area in the region of the Ukrainian Antarctic Station. The symbols $\Delta F$ and $\Delta \Delta F$ denote the difference of magnetic field at a given and base point for a particular year, and the change in the difference of magnetic field for a given and base point in a measured cycle, respectively. $C$ denotes average annual $^{222}Rn$ concentration at the earth surface}
\label{tab2}
\begin{tabular}{|c|c|c|c|c|c|c|c|c|c|c|c|c|}
\hline
\cellcolor[rgb]{0.55,0.7,0.89} \tiny{Location} & \tiny{Distance} & \multicolumn{4}{|c|}{\tiny{$\Delta F$, (nT)}} & \tiny{$\Delta \Delta F$} & \tiny{$C$} & \tiny{$\Delta \Delta F$} & \cellcolor[rgb]{0.55,0.7,0.89} \tiny{$C$} & \multicolumn{3}{|c|}{\tiny{$\Delta \Delta F$}, (nT)} \\

\cline{3-6} \cline{11-13}

\cellcolor[rgb]{0.55,0.7,0.89} \tiny{No. (n)} & \tiny{(km)} & & & \cellcolor[rgb]{0.55,0.7,0.89} & & \tiny{(nT)} & \tiny{($Bq ~ m^{-3}$)} & \tiny{(nT)} &   \cellcolor[rgb]{0.55,0.7,0.89} \tiny{($Bq ~ m^{-3}$)}  &  &  & \\

\cellcolor[rgb]{0.55,0.7,0.89} & & \tiny{2002} & \tiny{2003} &\cellcolor[rgb]{0.55,0.7,0.89} \tiny{2004} &  \tiny{2005} & \tiny{2003-02} & \tiny{2002-03} & \tiny{2004-03} & \cellcolor[rgb]{0.55,0.7,0.89} \tiny{2003-04} & \tiny{2005-02} & \tiny{2005-03} & \tiny{2005-02} \\
 
\hline
\cellcolor[rgb]{0.78,0.85,0.95} \tiny{1}   & \tiny{0}      & \tiny{289.2} & \tiny{292.1}   & \cellcolor[rgb]{0.78,0.85,0.95} \tiny{287.3}   & \tiny{289.4}   & \tiny{2.9}  &                  & \tiny{-4.8} & \cellcolor[rgb]{0.78,0.85,0.95} & \tiny{0.2}  & \tiny{-2.7} & \tiny{2.1} \\
\hline
\cellcolor[rgb]{0.78,0.85,0.95} \tiny{1a} & \tiny{0.3}   & \tiny{328.6} & \tiny{326.7}   & \cellcolor[rgb]{0.78,0.85,0.95} \tiny{325.7}   & \tiny{321.6}   & \tiny{-1.9} &                  & \tiny{-1}    & \cellcolor[rgb]{0.78,0.85,0.95} \tiny{34.7$\pm$6.6} & \tiny{-7.0} & \tiny{-5.1} & \tiny{-4.1} \\
\hline
\cellcolor[rgb]{0.78,0.85,0.95} \tiny{2}   & \tiny{0.8}   & \tiny{125.0} & \tiny{127.6}   & \cellcolor[rgb]{0.78,0.85,0.95} \tiny{126.6}   & \tiny{123.0}   & \tiny{2.6}  & \tiny{8.0} & \tiny{-1}    & \cellcolor[rgb]{0.78,0.85,0.95} \tiny{15.8$\pm$2.9} & \tiny{-2.0} & \tiny{-4.6} & \tiny{-3.6} \\
\hline
\cellcolor[rgb]{0.78,0.85,0.95} \tiny{3}   & \tiny{1.7}   & \tiny{567.7} & \tiny{570.9}   & \cellcolor[rgb]{0.78,0.85,0.95} \tiny{564.6}   & \tiny{562.9}   & \tiny{3.2}  & \tiny{8.7} & \tiny{-6.3} & \cellcolor[rgb]{0.78,0.85,0.95} \tiny{5.7$\pm$3.2}   & \tiny{-4.8} & \tiny{-8.0} & \tiny{-1.7} \\
\hline
\cellcolor[rgb]{0.78,0.85,0.95} \tiny{3a} & \tiny{1.9}   & \tiny{531.2} & \tiny{535.4}   & \cellcolor[rgb]{0.78,0.85,0.95} \tiny{531.5}   & \tiny{526.6}   & \tiny{4.2}  &                  & \tiny{-3.9} & \cellcolor[rgb]{0.78,0.85,0.95} & \tiny{-4.6} & \tiny{-8.8} & \tiny{-4.9} \\
\hline
\cellcolor[rgb]{0.78,0.85,0.95} \tiny{BP} & \tiny{2.9}   & \tiny{0}        & \tiny{0}          & \cellcolor[rgb]{0.78,0.85,0.95} \tiny{0}          & \tiny{0}          & \tiny{0}     & \tiny{4.0} & \tiny{0}     & \cellcolor[rgb]{0.78,0.85,0.95} \tiny{8.3$\pm$2.2}   & \tiny{0}     & \tiny{0}     & \tiny{0}     \\
\hline
\cellcolor[rgb]{0.78,0.85,0.95} \tiny{5}   & \tiny{3.3}   & \tiny{40.4}   & \tiny{41.5}     & \cellcolor[rgb]{0.78,0.85,0.95} \tiny{36.6}     & \tiny{38.7}     & \tiny{1.1}  &                  & \tiny{-4.9} & \cellcolor[rgb]{0.78,0.85,0.95} & \tiny{-1.7} & \tiny{-2.8} & \tiny{2.1}  \\
\hline
\cellcolor[rgb]{0.78,0.85,0.95} \tiny{4}   & \tiny{3.7}   & \tiny{77.6}   & \tiny{77.2}     & \cellcolor[rgb]{0.78,0.85,0.95} \tiny{75.7}     & \tiny{77.5}     & \tiny{-0.4} & \tiny{4.3} & \tiny{-1.5} & \cellcolor[rgb]{0.78,0.85,0.95} \tiny{20.9$\pm$4.8} & \tiny{-0.1} & \tiny{0.3}  & \tiny{1.8}   \\
\hline
\cellcolor[rgb]{0.78,0.85,0.95} \tiny{6}   & \tiny{7.0}   & \tiny{277.0} & \tiny{279.4}   & \cellcolor[rgb]{0.78,0.85,0.95} \tiny{278.2}   & \tiny{279.7}   & \tiny{2.4}  & \tiny{5.7} & \tiny{-1.2} & \cellcolor[rgb]{0.78,0.85,0.95} \tiny{10.7$\pm$3.4} & \tiny{2.7}  & \tiny{0.3}  & \tiny{1.5}   \\
\hline
\cellcolor[rgb]{0.78,0.85,0.95} \tiny{7}   & \tiny{10.5} &                      & \tiny{1015.9} & \cellcolor[rgb]{0.78,0.85,0.95} \tiny{1019.2} & \tiny{1028.8} &                   &                  & \tiny{3.3}  & \cellcolor[rgb]{0.78,0.85,0.95} \tiny{11.2$\pm$4.5} &  & \tiny{12.9} & \tiny{9.6} \\
 \hline
\end{tabular}
\end{center}
\end{table}

The geomagnetic observations offered information on the structure of the anomalous magnetic field $\Delta F$ along the profile I – I, and also its time changes – tectonomagnetic anomalies – $\Delta \Delta F$. It was found that the $\Delta F$ is strongly dependent on position (Table~\ref{tab2}). Due to discrepancy of the magnetic properties of volcanic and intrusive rocks the differences of magnetic anomaly were more then $1000~nT$ with the maximum in the eastern part of the profile, at Rasmussen Cape (point 7).  In the western part of the profile, on Three Little Pigs Islands (point 3), a local anomaly $\Delta F$ of intensity $600~nT$ was detected also. Dynamic changes of magnetic field $\Delta \Delta F _ {2003 - 02}$ during the period 2002 - 2003 (Table~\ref{tab2}) varied from $-1.9$ (Barchans Islands) to $+4.2~nT$ (Three Little Pigs Islands). The definite regularity of their distribution along the profile was noticeable despite the small value of $\Delta \Delta F$. During the periods 2003-2004 Barchans Islands and 2004-2005 these values amount $-6.3~nT$ (Three Little Pigs Islands) to $+3.3~nT$ (Barchans Islands) and $-4.9~nT$ (Three Little Pigs Islands) to $+6.9~nT$ (Rasmusen Cape). The repeated observation of 2003, 2004, and 2005 confirms, in general, the stability of $\Delta \Delta F$ distribution along the profile I – I observed in 2002 (Table~\ref{tab2}). Similar anomalous effects, but with somewhat smaller amplitude, are the general characteristic of active breaks in the Earth's crust and were revealed in many seismoactive zones of the world \citep{Maksymchuk2002}.

\subsection{Radon monitoring}

Long term measurements of radon concentration where carried out by passive radon dosimeter utilizing CR-39 and Makrofol E detectors. For this purpose the radon monitoring devices of the J. Stefan Institute (IJS), Ljubljana \citep{Sutej1986} and Nuclear Center Karlsruhe (KfK), Karlsruhe \citep{Urban1981} were used. The dosimeters were placed on the Earth's surface (with the opening of the dosimeter, covered by a filter, looking downwards) and/or about 40 cm above the Earth surface (the dosimeters were fixed on styropore support). The measurements performed in 2002-2003 by IJS dosimeter are described in \citep{Ilic2005}. Due to technical problems all of the measurements in 2003-2004 unfortunately were lost. At each sites at least 7 dosimeters were positioned to get information on the reproducibility of the results. The dosimeters were covered by a carbon steel oil drum (80~cm dia and 1~m in height, used for 200~L volume). The detectors were exposed about one year. Before and/or after the exposure the detector foils were placed in aluminized foil to eliminate Rn exposure during transport to/from Europe/Antarctica. The CR-39 foils were etched in 6.25~N~NaOH at~70$^{\circ}$C for 8~hours, and evaluated by the TRACOS system \citep{Skvarc2001}. The response of the dosimeter to $^{222}Rn$ is $0.15~(tracks \cdot cm^{-2}) / (Bq \cdot m^{-3} \cdot day)$. The background track density was about $100 ~ tracks / cm^2$. Etching and reading of KfK dosimeteres were performed in Karlsruhe. It should be mentioned that due to moisture collections KfK dosimeters are not able to measure the radon concentration at the surface of the Earth. Average results of radon concentration measurements are given in Table~\ref{tab2}.

\section{Results}
\label{sec4}

\begin{figure}
\begin{center}
\includegraphics[width=0.7\linewidth]{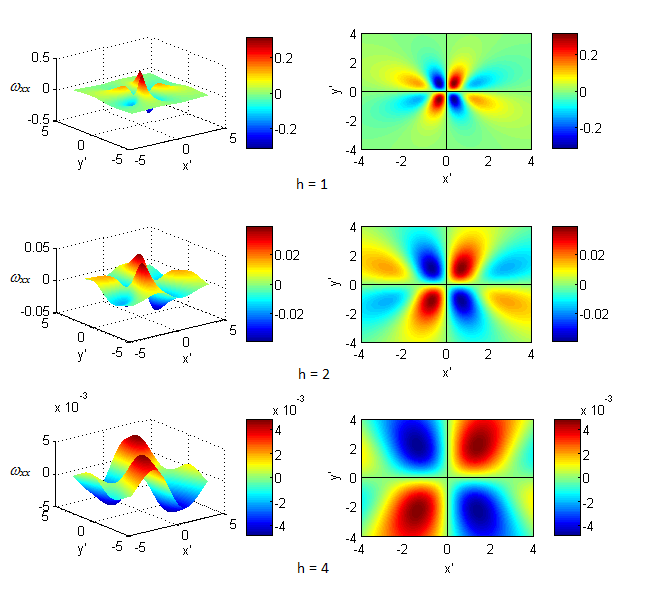}
\caption{The distribution of conditional deformation $\omega _{xx}$ at different value of $h$.}
\label{fig3}
\end{center}
\end{figure}

The isolines of conditional deformation $\omega _{xx}$ calculated by Eqs.~(\ref{eq11}) and~(\ref{eq15}) at Poisson's ratio $\nu = 0.25$ and the average depth of local heterogeneity occurrence $h = 1, 2, 4$ are shown in Fig.~\ref{fig3}. Note that symbol $h = H / \rho$~(\ref{eq14}). From Fig.~\ref{fig3} it is possible to see that the strain distribution is strongly irregular. For deformation $\omega _{xx}$ the coordinate axes, which coincide with the direction of maximal tangent stress in given region, are zero isolines and moreover there is an additional zero isoline in every quadrant. Thus, it is possible to miss the signal even quite close to the epicenter of tectonic activity, if the observation point is poorly chosen.

\begin{figure}
\begin{center}
\includegraphics[width=0.4\linewidth]{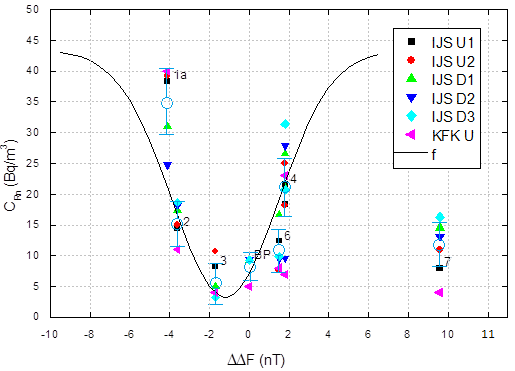}
\caption{Annual (2004-05) $^{222}Rn$ concentration, $C$, vs. change of diference of magnetic field, $\Delta \Delta F$ in the cycle 2002-03 at different locations 1a-7 (see Fig.~\ref{fig2}). IJS and KFK denote radon dosimeter of J. Stefan Institute, Ljubljana and Nuclear Research Center, Karlsruhe respectively. With D and U radon measurements at the Earth surface and 40 cm above the surface are denoted. $f$ is function $C_n = C_{BP} + C_{\infty} \left[ 1 - exp \left( - \left \vert \Delta \Delta F_n + \Delta \Delta F_d \right \vert ^2 / \left( \Delta \Delta F_0 \right)^2 \right) \right]$, which was fitted to the experimental data.}
\label{fig4}
\end{center}
\end{figure}

Experimental results of the annual radon concentration vs. change of difference of magnetic field for 2004-2005 are shown in Fig.~\ref{fig4}. Comparison of the radon concentration values and the tectonomagnetic results shows a clear correlation (see Table~\ref{tab3}). 

The following function was fitted to the experimental data (Fig.~\ref{fig4})

\begin{equation}
C_n = C_{BP} + C_{\infty} \left[ 1 - exp \left( - \left \vert \Delta \Delta F_n + \Delta \Delta F_d \right \vert ^2 / \left( \Delta \Delta F_0 \right)^2 \right) \right]
\label{eq37}
\end{equation}

\noindent where $C_{\infty}$ is the maximal radon concentration for a given area and $\Delta \Delta F_0$ is the change of the magnetic field, which corresponds to destructive (failure) stress, and $\Delta \Delta F_d$ is a constant describing displacement of the base point. It was found that $\Delta \Delta F_0 = 3.78 ~nT$, $C_{BP} = 3.3 ~ Bq / m^3$, $C_{\infty} = 5 ~Bq / m^3$ and $\Delta \Delta F_d = 1.2 ~nT$. Note when $\left \vert \Delta \Delta F_n \right \vert \ll \Delta \Delta F_0$, and $\Delta \Delta F_d  = 0$, the Eq.~(\ref{eq37}) is transformed to

\begin{equation}
C_n = C_{BP} + c \left \vert \Delta \Delta F_n \right \vert
\label{eq38}
\end{equation}

\noindent where $c = C_{\infty} / \Delta \Delta F_0$ was estimated to be of the order of magnitude $1~Bq~m^{-3} / nT$.

\begin{figure}
\begin{center}
\includegraphics[width=0.5\linewidth]{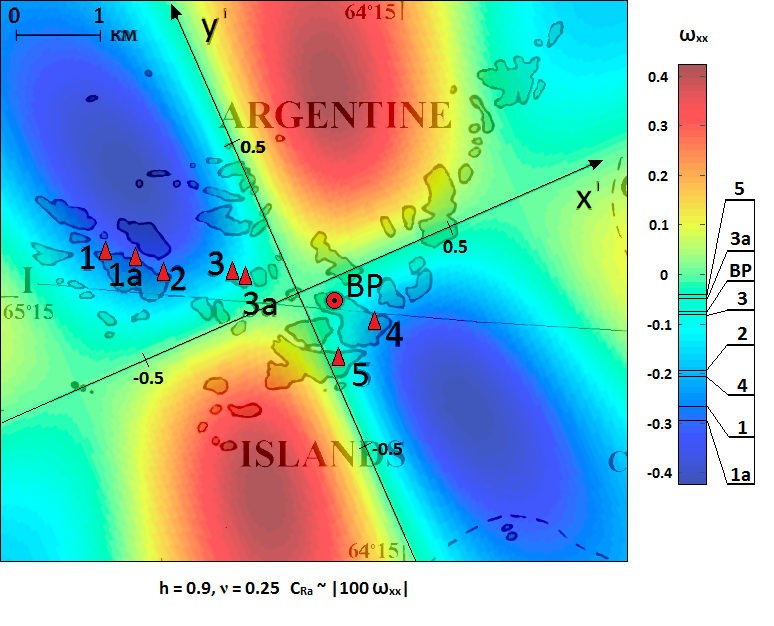}
\caption{The distribution of conditional deformation $\omega_{xx}$ on the Earth's daylight at displacement value of $h = 0.9$. Points 1, 1a, 2, 3, BP, 4, 5 are radon measurement points (see Fig.~\ref{fig2}). $C_{Rn}$, $\omega_{xx}$ and $\nu$ denote annual radon concentration, conditional deformation and Poisson's ratio, respectively.}
\label{fig5}
\end{center}
\end{figure}

Analysis of Figs.~\ref{fig3} and~\ref{fig4} (describing cross-correlation between two precursors – radon and magnetic field variations) shows that the coordinate centre (0, 0, 1) is situated near the measurement point No.~3 (Fig.~\ref{fig2}) corresponding to minimal radon and magnetic field variations. Basing on this analysis and magnetic measurement data \citep{Maksymchuk2003}, we plotted coordinates in Fig.~\ref{fig2}. The axis $x'$ coincides with the maximal tangent stress direction in the given region. The axis $y'$ traverses point (0, 0, 0.9) close to measurement point No. 3. After that we plot the coordinates of all measurement points on conditional deformation field at $h = 0.9$ (Fig.~\ref{fig5}). It is found that the radon concentration ratio for the different pairs of observation points is approximately equal to conditional deformation ratio for corresponding observation points, which is shown in Fig.~\ref{fig5}. Hence we may suppose that the spatial distributions of radon concentration and conditional deformation have linear dependence, but according to the commentary at the end of Section~\ref{sec2.1}, only within the extensional areas  (e.g. blue region on the Fig.~\ref{fig3} and Fig.~\ref{fig5})

\begin{equation}
C_{Rn} \sim 
\begin{cases}
-100 \omega_{xx}, & if ~~ \omega_{xx} < 0; \\
0, & if ~~ \omega_{xx} \geqslant 0.
\end{cases}
\label{eq39}
\end{equation}

It is interesting to note here that both the physically qualitative and quantitative senses are in good agreement with the values of radon background concentrations as measured at the points~6 and~7 (see. Fig.~\ref{fig2}, Table~\ref{tab2} and Fig.~\ref{fig4}), which correspond to the compressional areas (e.g. red region on the Fig.~\ref{fig3} and Fig.~\ref{fig5}).

Such result is in accordance with predictions of so-called dilatation models \citep{Kasahara1981}, which well explain observed seismic precursors (in spite of the fact that main of them, i.e. filtration (Scholz et~al., 1973) and "dry" \citep{Mjachkin1975} models are based on the fundamentally different physical processes).

This is a very important observation that should be thoroughly tested in further experiments. If this statement is correct, it can become a key moment for the understanding of the physical mechanism of radon emanation at increased tectonic activity, which, as it is well known, can disappear due to slow relaxation processes or fast destruction, i.e. earthquakes \citep{Durrani1997}. In our opinion, one of the convincing methods of the verification of the nonrandom existence of stated physical connection (a linear dependence between spatial radon distribution and conditional deformation) is the simultaneous identification of the points of observation relative to spatial distribution of magnetic field variations.

Since we make comparison with the field observation data in the region of the Western part of the Antarctica Peninsula, let us carry out preliminary analysis of Eq.~(\ref{eq35}) as applied to this region. Taking into account the distribution of background tectonic stress in this region \citep{Maksymchuk2003}, we direct the axes $x_2$ and $x_1$ as it shown in Fig.~\ref{fig2} and axis $x_3$ -- to the Earth's center. The absolute value of geomagnetic-field vector in the Western part of the Antarctica Peninsula is about 39500~nT, the inclination is $\theta _1 = -57 ^{\circ} 14 '$ and the declination is $\theta _2 = -16 ^{\circ} 37 '$. 

Let us place the center of area $V$ on axis $x_3$ as deep as $H \left( \xi_1 ^0 = \xi_2 ^0 = 0, \xi_3 ^0 = H \right)$ and pass on to dimensionless quantities

\begin{equation}
x = x_1 / \rho, ~~~ y = x_2 / \rho , ~~~ h = H / \rho , ~~~ r_d = r / \rho = \sqrt{x^2 + y^2 + \left( h - 0.5 \right) ^2}, 
\label{eq40}
\end{equation}

\begin{equation}
\beta_1 = \frac{2}{3} \frac{\rho^2 \Delta F_{r1}}{k C V \alpha I \tau} = \frac{1}{2 \pi} \frac{\Delta F_{r1}}{k C \alpha I \tau }.
\label{eq41}
\end{equation}

As a result of this Eq.~(\ref{eq35}) takes the form

\begin{equation}
\beta_1 = \frac{3}{2 r_d ^5} \left[ 2 ( b_1 ^2 + b_2 ^2 ) xy + b_2 b_3 x (2h - 1) + b_1 b_e y (2h - 1) \right],
\label{eq42}
\end{equation}

\noindent where $b_1 = \sin {(57 ^{\circ} 14 ')}$, $b_2 = \cos {(16 ^{\circ} 37 ')}$, $b_3 = \sin {(16 ^{\circ} 37 ')}$.

\begin{figure}
\begin{center}
\includegraphics[width=0.7\linewidth]{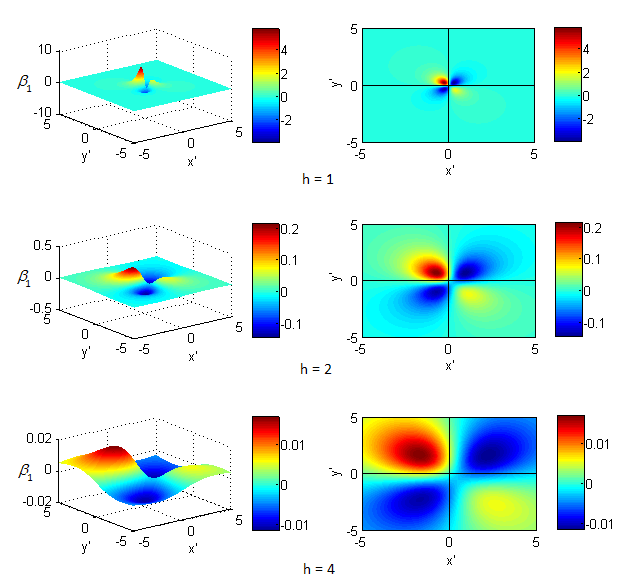}
\caption{Isolines of dimensionless magnetic induction field at different h on the Earth's daylight.}
\label{fig6}
\end{center}
\end{figure}

The isolines of dimensionless magnetic induction field $\beta _1$ at $h = 1, 3, 4$ are presented in Fig.~\ref{fig6}. 

This field has a complex structure dividing into 4 parts by orthogonal zero lines. At first glance, here are possible experimental situations as in the case of deformation or radon measurements (Fig.~\ref{fig2}), when the signal is not detected even in immediate proximity to the epicentre of tectonic activity, if observer's location is unsuccessful. But this is not the case. Below we will consider this situation and will show that it is much difficult and interesting in respect to observer. For that we consider the second item $\Delta F_{r2}$ of Eq.~(\ref{eq28})

\begin{equation}
\Delta F_{r2} = - \frac{3}{2} k C b_p e_{mkp} \nabla _m \int \int \int \frac{e_{ijk} I_l s_{il} '' R_j}{R^3} dv .
\label{eq43}
\end{equation}

Eqs.~(\ref{eq29}) and~(\ref{eq43}), which describe the first $\Delta F_{r1}$ and the second $\Delta F_{r2}$ components of the total variation of magnetic field $\Delta F_{r}$, have distinction of kind. In Eq.~(\ref{eq43}) the integration is over the all volume of $V$, therefore at sufficiently large distances from area $V$, Eq.~(\ref{eq29}) can be approximated by the field of equivalent magnetic dipole and decreases according to $1 / r^3$ law. As against Eq.~(\ref{eq29}) the integration in Eq.~(\ref{eq43}) is over the all volume of regular medium (including the vicinity of the point of observation on the Earth's daylight). The integral in Eq.~(\ref{eq43}) has dependence like $1 / r^2$ at $R \rightarrow 0$, hence, it is possible to expect that the main contribution to field $\Delta F_{r2}$ is made by medium volumes in the immediate region of observation point. According to \citet{Dobrovolsky1984}, such a property of the future solution makes it possible to reduce Eq.~(\ref{eq43}) to the following approximation

\begin{equation}
\Delta F_{r2} \simeq - 4 \pi k C I b_i b_l s_{il} '' .
\label{eq44}
\end{equation}

The independence of $\Delta F_{r2}$ on radius of the distinguished unperturbed medium following from Eq.~(\ref{eq44}) \citep{Dobrovolsky1984} is an important result. Hence, in homogeneous fields of stress, the magnetization $\Delta F_{r2}$ is determined by local effect according to Eq.~(\ref{eq44}). This means that, when Eq.~(\ref{eq44}) characterizes the magnetic field variations of unperturbed medium close to observation point, the addition $\Delta F_{r2}$ (specific for given region) may be found by the averaging of the sample of measurements, which made exceptionally on "zero" isolines of the spatial distribution of first component variations $\Delta F_{r1}$. In other words, the observation strategy for observation points location must not only take into account the real existence of "zero" isolines $\Delta F_{r1}$ but also to use them for the determination of average local addition $\Delta F_{r2}$. 

\begin{figure}
\begin{center}
\includegraphics[width=0.5\linewidth]{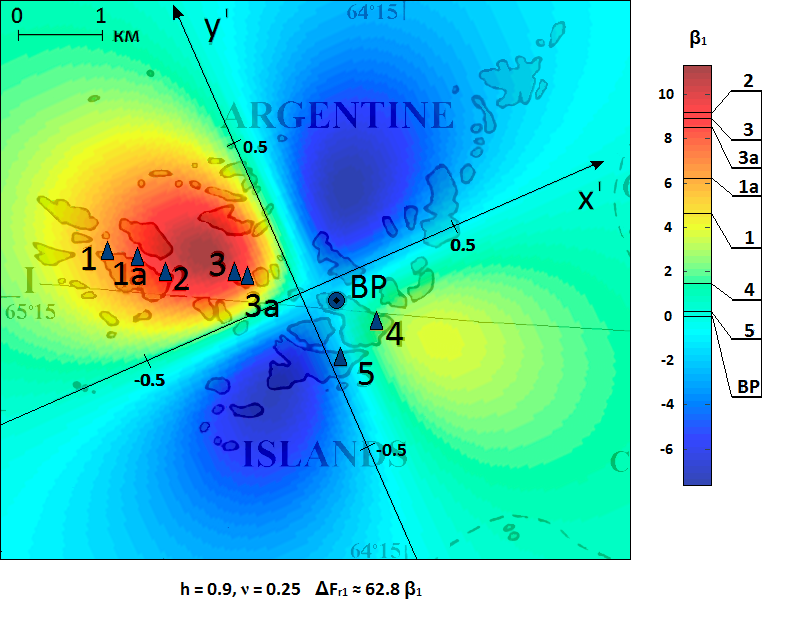}
\caption{The distribution of dimensionless magnetic induction field at displacement value $h = 0.9$. Points 1, 1a, 2, 3, BP, 4, 5 are the measurement points of magnetic field variations. Symbols $\Delta F_{r1}$, $\beta$ and $\nu$ denote the first component of total variation of magnetic field, dimensionless magnetic induction field and Poison's ratio respectively.}
\label{fig7}
\end{center}
\end{figure}

At last, using coordinate system presented in Fig.~\ref{fig2} let us plot coordinates of all measurement points on the field of magnetic induction increment $\Delta F_{r1}$ at $h = 0.9$ (Fig.~\ref{fig7}). To ground the choice of value $h = 0.9$ we compared the theoretical values $\Delta F_{r1}$ with the results of field observations in the Western part of the Antarctica Peninsula (Table~\ref{tab2}). To obtain the theoretical value of $\Delta F_{r1}$ (Eq.~(\ref{eq41})) we used such numerical values of parameters  \citep{Dobrovolsky1984}: $\alpha = 0.1$, $\tau = 100 ~MPa$, $k = 10^{-7} ~mT ~A^{-1}$, $CI = 10^{-8} ~A ~Pa^{-1} ~ m^{-1}$. Then we have

\begin{equation}
\Delta F_{r1} \approx 62.8 ~\beta_1, ~~nT.
\label{eq45}
\end{equation}

It is obvious that the theoretical value of $\Delta F_{r1}$ (Eq.~(\ref{eq45})) is equal to the experimental value (Table~\ref{tab2}), when $\beta _1 \sim 10$. This corresponds to distribution of magnetic induction increment $\Delta F_{r1}$ on the Earth's daylight just at $h = 0.9$ (Fig.~\ref{fig7}). 

In that case, if we neglect the second component $\Delta F_{r2}$, the ratio of measured variations of magnetic field for the different pairs of observation point (Fig.~\ref{fig2}) becomes approximately equal to the ratio of the theoretical values of magnetic field variations at the corresponding points of observation (Fig.~\ref{fig7}). At the same time the experimental value of first components $\Delta F_{r1}$ in point No 2 (Table~\ref{tab2}) is substantially smaller then its theoretical value (see Fig.~\ref{fig7}). This is explained by the fact that in this case it is not allowed to neglect second component $\Delta F_{r2}$, which (as it was mentioned above) is the characteristic of the piezomagnetic properties of medium in the immediate region of observation point. In other words, the total value of additional magnetic field $\Delta F_{r1}$ in the point No.2 decreases by $\Delta F_{r2}$ due to the strong (anomalous) piezomagnetic effect of medium in the point No.2 (the east part of Barchans Island).

\begin{figure}
\begin{center}
\includegraphics[width=0.5\linewidth]{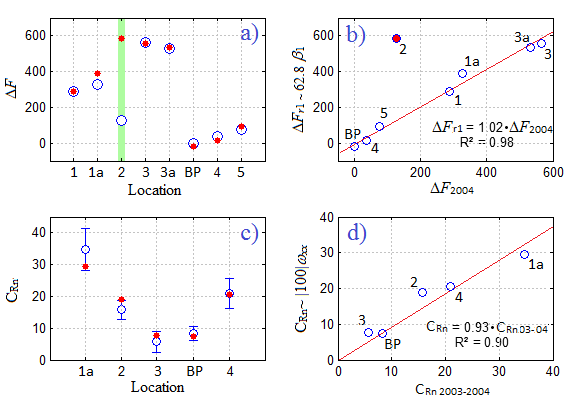}
\caption{Comparison between the experimental (blue circles with bars) and theoretical (red points) data for the variations of (a) — magnetic field and (c) – radon concentrations. Experimental evidence of the linear dependence of (b) – magnetic field variations on magnetic induction $\beta _1$~(\ref{eq45}) and (d) – radon concentrations on the conditional deformation $\omega _{xx}$~(\ref{eq39}). Note: $R^2$ is a coefficient of determination; anomalous behavior of the point 2 (green line) on Fig.~\ref{fig8}a is discussed in the text.}
\label{fig8}
\end{center}
\end{figure}

So, we may conclude that the selected strategy of measurements (the dashed line of the simultaneous measurements of radon concentration and magnetic field variations in Fig.~\ref{fig2} and Table~\ref{tab3}) should lead to the strong cross-correlation (Fig.~\ref{fig8}) between the spatial distributions of the value $\Delta F_{r}$ variations (i.e., $\Delta \Delta F_{r}$) and radon variations in all points of observation (except for points 6 and 7, which are, apparently, the characteristics of totally other physical localization of tectonic activity).

\begin{table}
\begin{center}
\caption{Results of geomagnetic field and radon concentration measurements at the tectonomagnetic area in the region of the Ukrainian Antarctic Station.}
\label{tab3}
\begin{tabular}{|c|c|c|c|c|c|c|}
\hline
Location & $\Delta F_{2004}$& \multirow{2}{*}{$\beta_1$} & Theory & $C_{2003-04}$ &  \multirow{2}{*}{$\omega_{xx}$} & Theory \\
No. (n) & (nT) &  & (nT) & ($Bq ~ m^{-3}$) &  & ($Bq ~ m^{-3}$) \\
\hline
1 & 287.3 & 4.6 & 288.9 & - & -0.275 & 27.5 \\ 
1a & 325.7 & 6.2 & 389.4 & 34.7$\pm$6.6 & -0.295 & 29.5 \\
2 & 126.6 & 9.3 & 584.0 & 15.8$\pm$2.9 & -0.190 & 19.0 \\
3 & 564.6 & 8.9 & 558.9 & 5.7$\pm$3.2 & -0.078 & 7.8 \\
3a & 531.5 & 8.5 & 533.8 & - & -0.050 & 5.0 \\
BP & 0 & 0 & 0 & 8.3$\pm$2.2 & -0.075 & 7.5 \\
5 & 36.6 & 0.2 & 12.6 & - & -0.040 & 4.0 \\
4 & 75.7 & 1.5 & 94.2 & 20.9$\pm$4.8 & -0.205 & 20.5 \\
 \hline
\end{tabular}
\end{center}
\end{table}

In our opinion, a good coincidence of theoretical and experimental samples of magnetic field variations in this area and simultaneous cross-correlation of the spatial distributions of magnetic field variations $\Delta \Delta F_r$ and radon variations is a cogent argument in favor of the assumption of linear dependence between spatial distributions of radon and conditional deformation 

Finally, the comparative analysis of the spatial distribution of points of observation and their location in Figs.~\ref{fig5} and~\ref{fig7}, which describe the variation distribution of radon and magnetic field, shows that the average size $\rho$ of local inhomogeneity along the profile of magnetic measurements  is approximately equal to $\rho \sim 4.0 ~km$. We can suppose that this size is comparative to the radius of local inhomogeneity $V$. If we suppose in turn that the stress of this volume will discharge by destruction i.e. by the earthquake, then, knowing radius $\rho$ of local inhomogeneity $V$, it is easy to calculate the magnitude M of potential earthquake \citep{Dambara1966}

\begin{equation}
\lg  \rho ~ [km] ~ = 0.5 M - 2.27 \Rightarrow M \simeq 5.8
\label{eq46}
\end{equation}

\noindent and also to estimate the precursor time $T$ \citep{Whitcomb1973}

\begin{equation}
\lg T ~ [day] ~ = 0.8 M - 1.92 \Rightarrow T = 525 ~days.
\label{eq47}
\end{equation}

\begin{figure}
\begin{center}
\includegraphics[width=0.7\linewidth]{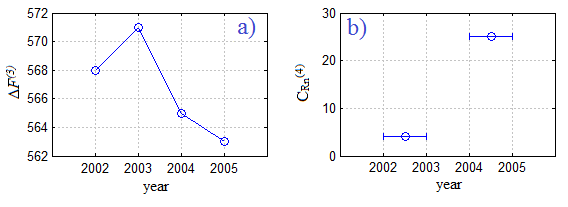}
\caption{The time dependences of the maximally observed variations of magnetic field $\Delta F$ (point 3) and radon concentration $C_{Rn}$ (point 4).}
\label{fig9}
\end{center}
\end{figure}

In order to determine an initial moment of precursor appearance let us plot the time dependences of maximally observed variations of magnetic field (point 3) and radon (point 4), in which the long-term parallel measurements were made. From the dependence shown in Fig.~\ref{fig8} follows that the initial moment of precursor appearance dates from the end of 2003 to beginning of 2004. Within the framework of dilatation models such behavior of precursors (Fig.~\ref{fig9}) can be explained by the occurrence of active stage of rock shattering (second stage of seismic cycle in the filtration \citep{Scholz1973} and "dry" \citep{Mjachkin1975} models), which can result in reduction of magnetic field $\Delta F_{r}$ and the increase of radon emanation, respectively.

\begin{figure}
\begin{center}
\includegraphics[width=0.5\linewidth]{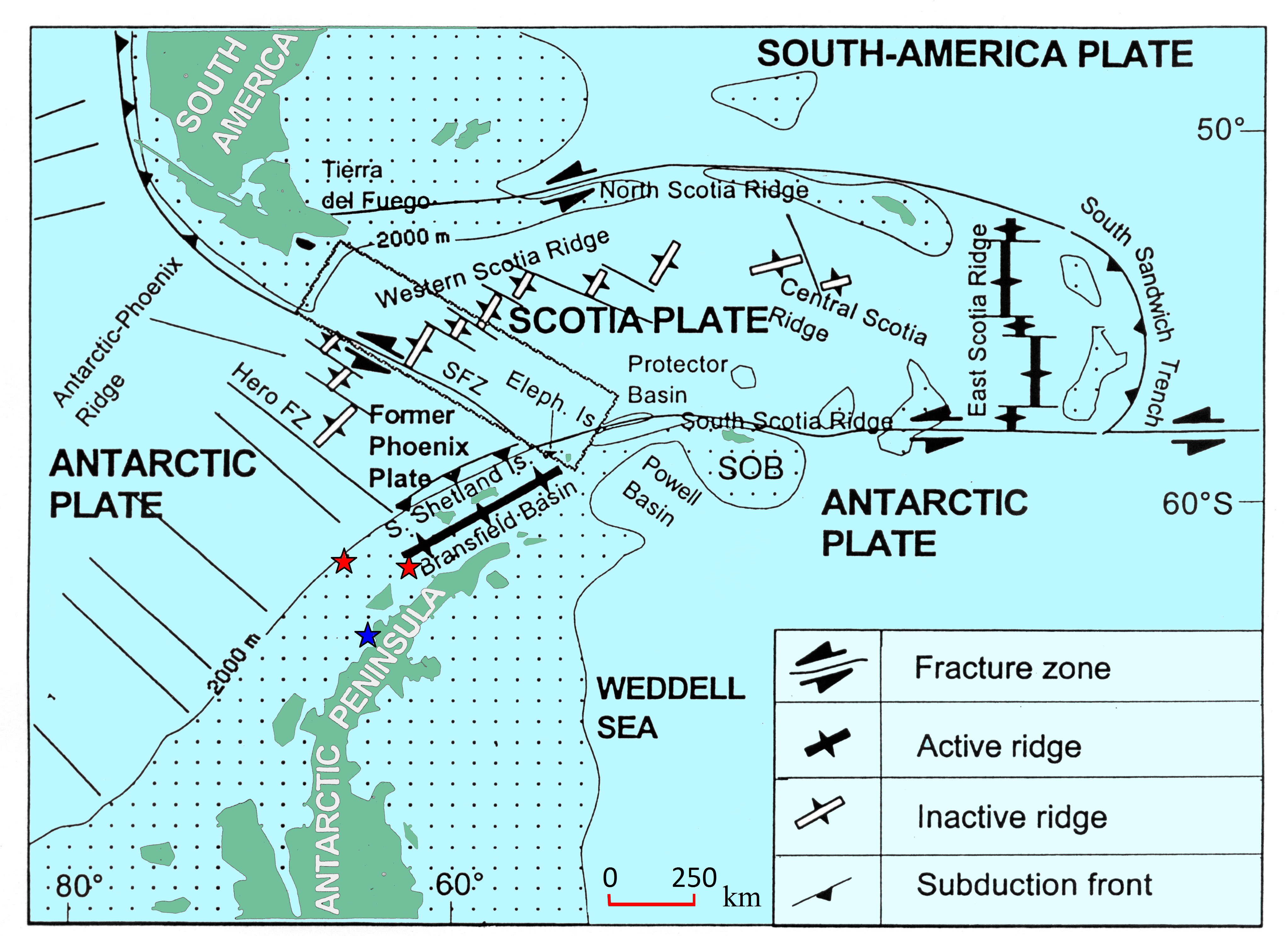}
\caption{Tectonic map of the boundary between the Scotia plate and Antarctic plate. Blue star marks the location of Ukrainian Antarctic station "Academician Vernadsky"(W64$^{\circ}$16$'$, S65$^{\circ}$15$'$), red stars mark the coordinates of the earthquakes with the magnitudes M5.4  (W61$^{\circ}$56$'$24$''$, S63$^{\circ}$28$'$48$''$) and M5.9 (W65$^{\circ}$49$'$48$''$, S62$^{\circ}$3$'$36$''$), which occurred in 2005 and 2007 respectively.}
\label{fig10}
\end{center}
\end{figure}

Thus, basing on Eq.~(\ref{eq47}) we could expect at the end of 2005 in this region the earthquake with epicenter nearby the Three Little Pigs island and magnitude $M = 5.8$. However, it did not happen because another two earthquakes with similar magnitudes occurred relatively near (at the distances of $\sim 250 ~km$ in 2005 and $\sim 320 ~km$ in 2007 (Fig.~\ref{fig10})) and apparently, "removed" the tectonic stress nearby the Three Little Pigs island.

Several remarks should be made. First, the stress of disturb area can be discharged not only by the destruction of the Earth crust \citep{Rusov2010}, but also by the slow mechanisms, for example, by the creep along the breaks. Other mechanisms are also possible, for example, dislocation sliding. If we take into account in addition that the region of Argentine Islands did not reveal seismic activity during the period of observation, then we can conclude that the earthquake in this region has a small probability.

Second, the empirical equations of the type~(\ref{eq46}), (\ref{eq47}) have the statistical character, and we can predict the day of the earthquake on its base only with some probability, if we even know the day of precursors appearance. Moreover, the constants in these equations, which was obtained by statistical averaging of numerous earthquakes, are different in different regions of the Earth, and are unknown for the considered region. 

\section{Conclusion}

Analysis of the theory and practices of the considered above parallel measurements of the spatial distribution of secondary precursor variations (magnetic field and radon) shows that the problem definition of the estimation of tectonic activity parameters must contain following strategy of measuring and data processing: 
\begin{enumerate}
\item In order to plot regional coordinates system, which characterizes the area of increased tectonic activity, it is necessary to determine the directions of maximal tangent stresses in given region.

\item Divide the given region into the grid with array pitch about $1 - 2 ~km$ and place measuring devices of radon and magnetic field variations in mesh points.

\item Using the symmetry properties of the cross-correlation of magnetic field and radon variations, determine the center of coordinate system (epicenter of tectonic activity).

\item Determine the absolute value, inclination and declination of magnetic-field vector in the investigated region.

\item Plot the theoretical spatial distribution of deformation and magnetic field variations (first component $\Delta F_{r1}$) considering region characteristics at different ($h = 1 \div 5$) depth of local heterogeneity occurrence (perturbed area $V$).

\item Plot on a regional coordinate system, with its center and axes selected according to item 3, the coordinates of all measurements of radon and magnetic field variations made according to item 2.

\item Make the optimal matching of regional coordinate system containing the distribution of experimental measurement point coordinates with the theoretical distribution of deformation (radon) and magnetic field (first component $\Delta F_{r1}$) variations plotted according to item 5.

\item In the case of optimal matching (see item 7) determine the coordinates and directions of "zero" isolines of the spatial distribution of first component $\Delta F_{r1}$ variations and make an additional set of measurements of second component $\Delta F_{r2}$.

\item Build the theoretical spatial distributions of total variations of magnetic field when $\Delta F_{r} = \Delta F_{r1} + \Delta F_{r2}$ and make more exact procedure of item 7.

\item Using the procedure of item 7 (specified in item 9) determine the average radius $\rho$ of local inhomogeneity (perturbed area $V$) and estimate the magnitude $M$ by approximate formula~(\ref{eq46}). 

\item For given magnitude $M$ estimate the precursor time $T$ of possible earthquake by Eq.~(\ref{eq47}) or by other known approximate formulas (see, for example, Kasahara's monograph \citeyearpar{Kasahara1981}).

\end{enumerate}

Making all procedures, it is necessary to take into account all remarks described at the end of previous section.

Turning back to the main results of this paper, one can make the following conclusions. The model idea of the precursors appearance mechanisms stimulated by tectonic activity and some peculiarities of the observer's strategy, aimed at the effective measurement of precursors within their spatial zone of manifestation in the day-side of the Earth, are studied. In particular, a Dobrovolsky approximation \citeyearpar{Dobrovolsky1984} is considered, which means that a non-perturbed medium (characterized by simple shear state) and a tectonic activity area (local inhomogeneity, induced by a change in the shear modulus only) are linearly elastic, while the perturbation itself (particularly, a surface shear $w_r$~(\ref{eq8})-(\ref{eq10})) may be calculated as a difference between the solutions of the two separate static problems in the framework of elasticity theory with the same boundary conditions at the surface $S$ (Fig.~\ref{fig1}).

This approximation let us derive a more accurate Dobrovolsky equation~(\ref{eq35}) for the spatial distribution of the magnetic field (first component) variations induced by the piezomagnetic effect under the perturbation of a regular medium in a state of a simple shear. Strong arguments in favor of the linear connection between the radon and conventional strain spatial patterns assumption are obtained.

Finally we can conclude the following. It has been shown theoretically that the deformations as a primary precursor may be a physical cause of the secondary precursors such as radon~(\ref{eq39}) and magnetic field~(\ref{eq45}). On the other hand, a verification of theoretical results obtained within a modified Dobrovolsky model based on the long-term radon and magnetic field variations measurement near the "Academician Vernadsky" antarctic station demonstrates a good agreement between the theoretical and experimental data (Fig.~\ref{fig5}, \ref{fig7} and~\ref{fig8}). A rather detailed strategy of measuring and data processing for the future experiments is presented as well.

The main conclusion we can make from the above discussion is as follows: studying the tectonic activity of the given region, it is necessary to use the complex approach measuring several known precursors and using some theoretical model. Only such complex approach can give the base to predict possible earthquake in the seismic regions. Concerning the region of Argentine Islands such approach permits to reveal the deformation accumulation in this region which will probably discharge by the slow relaxation mechanisms.

The cause of such accumulation is still unclear, and it should be analyzed using all data about the geological structure of the given region. In any case, the revealing of deformation accumulation in the given region should attract the additional attention to this area.

\section*{Acknowledgements}

This work is supported by EU FP7 Marie Curie Actions, SP3-People, IRSES project BlackSeaHazNet (PIRSES-GA-2009-246874).

\bibliographystyle{plainnat}

\bibliography{Rusov-StressTransferChanges}

\end{document}